\documentclass[fleqn,usenatbib]{mnras}

\usepackage{newtxtext,newtxmath}
\usepackage[T1]{fontenc}

\DeclareRobustCommand{\VAN}[3]{#2}
\let\VANthebibliography\thebibliography
\def\thebibliography{\DeclareRobustCommand{\VAN}[3]{##3}\VANthebibliography}

\newcommand{\rev}[1]{#1}

\usepackage{graphicx}	% Including figure files
\usepackage{amsmath}	% Advanced maths commands
\usepackage{mathrsfs}
\title[Momentum Resolution in Boltzmann Transport]{Momentum Space Resolution Dependence in Boltzmann Neutrino Radiation Hydrodynamics Simulations of Core-collapse Supernovae}

\author[R. Akaho]{Ryuichiro Akaho,$^{1}$\thanks{E-mail: akaho@asagi.waseda.jp}
\\
$^{1}$
Faculty of Science and Engineering, Waseda University, 3-4-1 Okubo, Shinjuku, Tokyo 169-8555, Japan\\}

\date{Accepted XXX. Received YYY; in original form ZZZ}

\pubyear{\the\year{}}

\begin{document}
\label{firstpage}
\pagerange{\pageref{firstpage}--\pageref{lastpage}}
\maketitle

% Abstract of the paper
\begin{abstract}
Finite momentum space resolution is a primary source of uncertainty in solving the Boltzmann equation with the discrete-ordinates method.
In this paper, the momentum space resolution dependence of Boltzmann neutrino transport is studied by performing a series of two-dimensional core-collapse supernova (CCSN) simulations.
The effects on the explosion dynamics are discussed by individually varying the resolutions of the zenith and azimuth angles in momentum space, and of the energy.
It is found that a coarse zenith angle resolution artificially facilitates the explosion, even turning non-exploding models into exploding ones.
This is because the coarse zenith angle grid cannot capture the forward-peaked distribution, thereby underestimating the flux factor and making neutrinos stay in the gain region for a longer time.
The dependence on the azimuth angle resolution is found to have little effect for the present non-rotating models, which is understandable given the sphericity of the CCSN core.
In contrast to the zenith angle resolution, a coarse energy resolution is found to artificially suppress the explosion.
This is because the neutrino heating rate at intermediate energies ($\sim30$--$50\,\mathrm{MeV}$) is underestimated in the low-resolution case.
Finally, the resolution dependence is studied for a one-dimensional black hole (BH) forming model. Unlike the explosion dynamics, BH formation does not exhibit a strong angular dependence. On the other hand, a coarse energy resolution is found to delay the BH formation time. This is because the momentum feedback is overestimated at low resolution, which allows a more massive neutron star to be supported.
\end{abstract}

% Select between one and six entries from the list of approved keywords.
% Don't make up new ones.
\begin{keywords}
supernovae: general -- neutrinos -- radiative transfer 
\end{keywords}

%%%%%%%%%%%%%%%%%%%%%%%%%%%%%%%%%%%%%%%%%%%%%%%%%%

%%%%%%%%%%%%%%%%% BODY OF PAPER %%%%%%%%%%%%%%%%%%

\section{Introduction}

Neutrinos are the main energy mediators in core-collapse supernova (CCSN) explosions (for reviews, see \citealt{Janka2017hsn..book.1095J,Janka2017hsn..book.1575J,Mezzacappa2020LRCA....6....4M,Burrows2021Natur.589...29B,Mezzacappa2023IAUS..362..215M,Boccioli2024Univ...10..148B,Yamada2024PJAB..100..190Y,Suzuki2024PTEP.2024eB101S,Janka2025ARNPS..75..425J}), and theoretical understanding hinges on the numerical treatment of neutrinos.
Neutrino transport is one of the most challenging, but also crucial, ingredients of CCSN simulations.
When the quantum nature of neutrinos (such as neutrino oscillations) is ignored, the neutrino transport in the CCSN core is described by classical kinetic theory \citep{Lindquist1966AnPhy..37..487L,Ehlers1971grc..conf....1E,Cercignani2002rbet.book.....C,Sarbach2014CQGra..31h5013S}, and the simulations require numerically solving the Boltzmann equation for the phase space distribution function.

Due to the daunting computational cost of the Boltzmann neutrino transport, the current standard approach in the literature is the truncated moment method \citep{Thorne1981MNRAS.194..439T,Shibata2011PThPh.125.1255S}, where the angular moments of the distribution function are treated as the key quantities.
However, the moment equation of a given order depends on the next higher order moment. Therefore, the closure relations are required to close the system \citep{Murchikova2017MNRAS.469.1725M}, and these are the main source of uncertainty in the truncated moment method (but see also \citealt{Wang2023ApJ...943...78W}).
In addition, the full momentum-angle information becomes very important for treating neutrino fast flavour conversion (FFC), which has recently been attracting great attention (for reviews, see \citealt{Tamborra2021ARNPS..71..165T,Capozzi2022Univ....8...94C,Richers2022arXiv220703561R,Volpe2024RvMP...96b5004V,Johns2025ARNPS..75..399J}).
The onset of FFC, the fast flavour instability (FFI), is known to be equivalent to ELN-XLN (electron minus heavy-lepton number) crossings in momentum space \citep{Morinaga2022PhRvD.105j1301M,Dasgupta2022PhRvL.128h1102D,Fiorillo2024JHEP...08..225F}.
Self-consistently incorporating the effects of FFC onto classical transport simulation requires the multiangle transport \citep{Xiong2025PhRvL.134e1003X,Akaho2026PhRvL.136s1002A}.
If the moment method is used instead, the limited angular information that it retains is insufficient to identify FFI, and phenomenological prescriptions such as the density threshold method \citep{Ehring2023PhRvD.107j3034E,Ehring2023PhRvL.131f1401E,Mori2025PASJ...77L...9M} must be relied upon, which also carry parameter uncertainty.
There have been several attempts to identify FFI from lower moments \citep{Nagakura2021PhRvD.103l3025N,Abbar2024PhRvD.109d3024A,Abbar2024PhRvD.109h3019A,Wang2025ApJ...986..153W,Wang2026ApJ...997..325W}. However, the accuracy of such methods is uncertain and they can have large systematic errors, such as underestimation of the crossings \citep{Cornelius2025PhRvD.112f3006C}.

There have been several attempts to solve the Boltzmann neutrino transport for CCSNe, using the discrete-ordinates ($S_N$) method \citep{Mezzacappa1993ApJ...405..669M,Yamada1999A&A...344..533Y,Liebendorfer2004ApJS..150..263L,Liebendorfer2005ApJ...620..840L,Livne2004ApJ...609..277L,Sumiyoshi2005ApJ...629..922S,Ott2008ApJ...685.1069O,Sumiyoshi2012ApJS..199...17S,Chan2020MNRAS.496.2000C} or the spectral method \citep{Peres2014CQGra..31d5012P}.
Monte Carlo neutrino transport is another multiangle method that can serve as an alternative \citep{Janka1991ntts.book.....J,Janka1992A&A...256..452J,Abdikamalov2012ApJ...755..111A,Kato2020ApJ...897...43K,Kato2021ApJS..257...55K}.
It is also worth mentioning that there is a growing interest in solving the Monte Carlo neutrino transport in neutron star merger and collapsar disk simulations \citep{Richers2015ApJ...813...38R,Miller2019PhRvD.100b3008M,Foucart2020ApJ...902L..27F,Kawaguchi2025PhRvD.111b3015K,Kawaguchi2025PhRvD.112d3001K}.

In the limit of infinite momentum space resolution, the Boltzmann transport provides a first-principles answer to the CCSN problem.
In practice, however, the computational cost of these transport methods remains quite high, making it infeasible to reach the resolution convergence.
When the Boltzmann equation is solved implicitly, which is a common practice when the stiff neutrino-matter interaction terms are treated, the computational cost scales as the square of the number of momentum space grid points.
Such a high cost restricts the momentum space resolution to a finite value, which constitutes a primary uncertainty when interpreting the results obtained with the Boltzmann transport.

The impact of the spatial resolution on the neutrino-driven mechanism and turbulence has been studied extensively \citep{Hanke2012ApJ...755..138H,Abdikamalov2015ApJ...808...70A,Radice2016ApJ...820...76R,Nagakura2019MNRAS.490.4622N}, and various codes have been compared in spherical symmetry \citep{OConnor2018JPhG...45j4001O}. In contrast, it remains poorly understood how the CCSN dynamics are affected by the momentum space resolution.
So far, the momentum angle resolution dependence has been examined only in 1D spherical symmetry \citep{Mezzacappa1993ApJ...405..637M,Thompson2003ApJ...592..434T,Liebendorfer2004ApJS..150..263L}, and the energy resolution dependence has been tested only in 1D with the Boltzmann transport \citep{Nakazato2007ApJ...666.1140N}, or in multiple dimensions with the moment method \citep{Marek2009ApJ...694..664M,Nagakura2021MNRAS.500..696N}.
Unlike the angular resolution, the energy resolution is a common ingredient of both the Boltzmann and the moment methods, so its dependence is of broad relevance \citep{OConnor2015ApJS..219...24O,Just2015MNRAS.453.3386J,Skinner2019ApJS..241....7S}.
Note that, although \citet{Richers2017ApJ...847..133R} carried out 2D Boltzmann calculations with different resolutions, these were restricted to steady-state configurations and varied the energy and $\theta_\nu$ resolutions simultaneously rather than one at a time.

In this paper, the momentum space resolution dependence of Boltzmann neutrino transport simulations is discussed by performing a series of two-dimensional (2D) Boltzmann neutrino radiation hydrodynamics simulations of CCSN explosions, together with 1D black hole (BH) forming simulations.
The resolution is varied for three meshes individually, namely the zenith and azimuth angles in momentum space, and the energy.
This is the first study of the momentum space resolution dependence of multi-dimensional Boltzmann neutrino transport in dynamical CCSN simulations.

This paper is organized as follows.
Section \ref{sec:setup} presents the numerical setup, covering the basic equations and the momentum space discretization.
The results are discussed in Section \ref{sec:results}, and Section \ref{sec:conclusion} closes the paper.
Natural units are employed throughout unless otherwise stated.

\section{Numerical Setup}
\label{sec:setup}

In this paper, the general relativistic Boltzmann neutrino radiation hydrodynamics code \citep{Nagakura2014ApJS..214...16N,Nagakura2017ApJS..229...42N,Nagakura2019ApJ...878..160N,Akaho2021ApJ...909..210A,Akaho2023ApJ...944...60A} is employed.
The basic equations are outlined in Section \ref{sec:equations}.
Since the focus of this paper is the momentum space resolution, the momentum space discretization methods are described in Section \ref{sec:discretize}.

\subsection{Basic Equations}
\label{sec:equations}
\begin{figure}
    \centering
    \includegraphics[width=\linewidth]{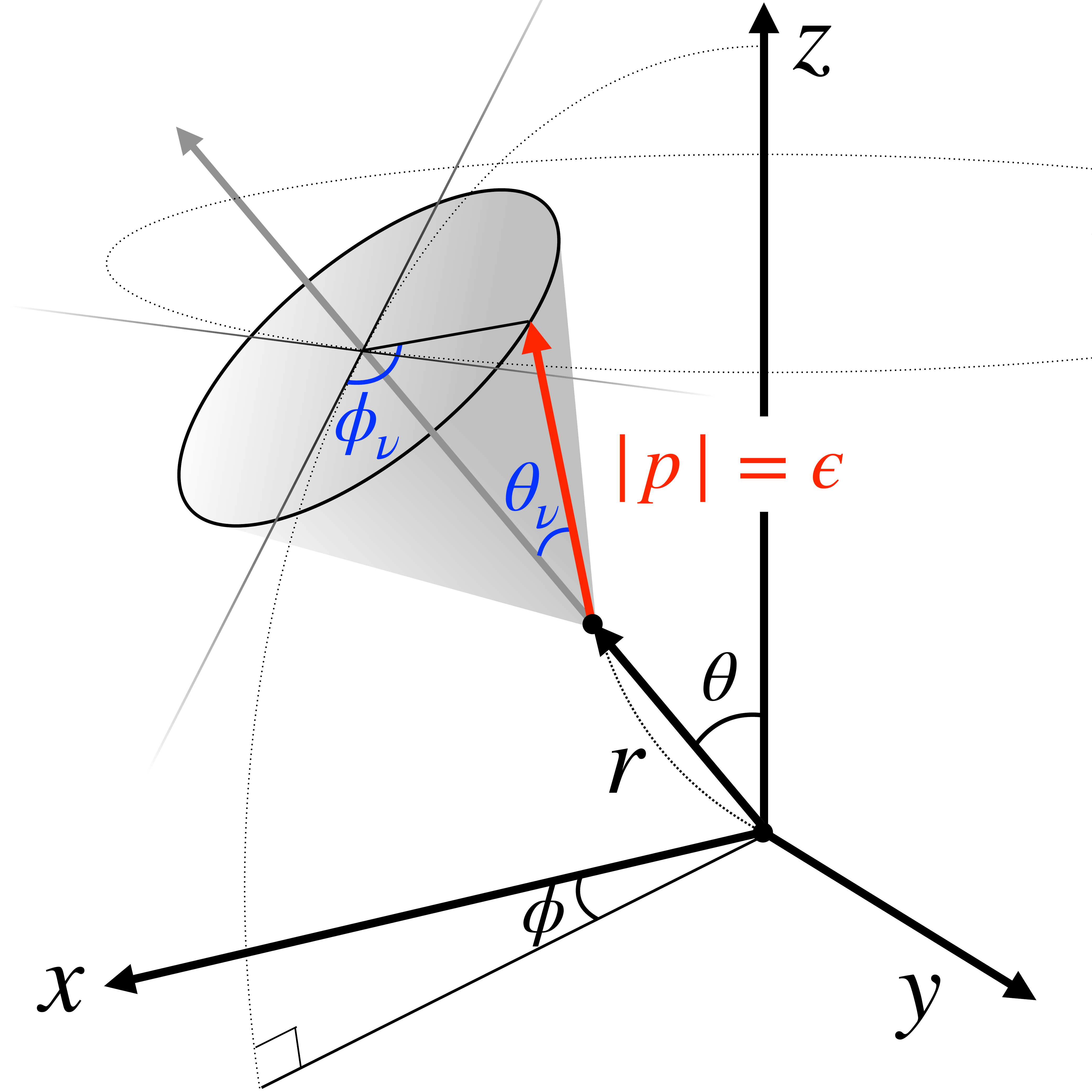}
    \caption{Schematic picture of the coordinate system used in this study. The red arrow represent the neutrino momentum.}
    \label{fig:coordinates}
\end{figure}
The code solves the Boltzmann equation with respect to the phase space distribution function $f$, written in the conservative form \citep{Shibata2014PhRvD..89h4073S}:
\begin{align}
\label{eq:boltz}
& \frac{1}{\sqrt{-g}}\frac{\partial}{\partial x^\mu}\left[\left(e_{(0)}^{\mu} + \sum_{i=1}^{3}l_i {e_{(i)}^\mu}\right)\sqrt{-g}f\right] 
-\frac{1}{\epsilon^2}\frac{\partial}{\partial\epsilon}\left(\epsilon^3f\omega_{(0)}\right) \nonumber \\
& +\frac{1}{\sin\theta_\nu}\frac{\partial}{\partial\theta_\nu}\left(\sin\theta_\nu f\omega_{(\theta_\nu)}\right) - \frac{1}{\sin^2\theta_\nu}\frac{\partial}{\partial{\phi_\nu}}\left(f\omega_{({\phi_\nu})}\right) = S_{\rm rad}.
\end{align}
Polar coordinates are used both for the configuration space and momentum space, as shown in Fig. \ref{fig:coordinates}.
As for the configuration space, $x^0=t$, $x^1=r$, $x^2=\theta$ and $x^3=\phi$ represent the time, radius, zenith and azimuth angles, respectively.
In momentum space, $\epsilon$ denotes the energy, $\theta_\nu$ and $\phi_\nu$ denote the zenith and azimuth angles, respectively.
In this paper, the spacetime geometry is limited to 2D, uniform with respect to $\phi$.
The directional cosines are given as
\begin{equation}
l_{(1)} = \mathrm{cos}\theta_\nu, \quad
l_{(2)} = \mathrm{sin}\theta_\nu\mathrm{cos}{\phi_\nu}, \quad 
l_{(3)} = \mathrm{sin}\theta_\nu\mathrm{sin}{\phi_\nu}. 
\end{equation}
The following quantities are introduced:
\begin{align}
& \omega_{(0)} \equiv \epsilon^{-2} p^\mu p_\nu \nabla_\mu e_{(0)}^\nu, 
\quad
\omega_{(\theta_\nu)} \equiv \sum_{i=1}^{3}\omega_{(i)}\frac{\partial l_{(i)}}{\partial \theta_\nu} \\
& \omega_{(\phi_\nu)} \equiv \sum_{i=2}^{3}\omega_{(i)}\frac{\partial l_{(i)}}{\partial \phi_\nu}, \quad
\omega_i \equiv \epsilon^{-2} p^\mu p_\nu \nabla_\mu e_{(i)}^\nu.
\end{align}
For the diagonal spacetime metric assumed in this study, the tetrad basis is chosen as
\begin{align}
& e^\mu_{(0)} = n^\mu \equiv \frac{1}{\alpha}\left(\frac{\partial}{\partial t}\right)^\mu, \quad
e^\mu_{(1)} = 
\frac{1}{\sqrt{\gamma_{rr}}}\left(\frac{\partial}{\partial r}\right)^\mu, \\
& e^\mu_{(2)} = \frac{1}{\sqrt{\gamma_{\theta\theta}}}
\left(\frac{\partial}{\partial\theta}\right)^\mu, \quad
e^\mu_{(3)} = \sqrt{\gamma^{\phi\phi}}
\left(\frac{\partial}{\partial \phi}\right)^\mu.
\end{align}
Here $n^\mu$ is the timelike unit normal to the spatial hypersurface, and $\alpha$, $\beta^i$, $\gamma_{ij}$ and $\gamma$ denote the lapse function, shift vector, spatial metric, and the determinant of it, respectively.
The right-hand side of the Boltzmann equation, $S_{\rm rad}$, denotes the collision terms.

The hydrodynamics equations are based on \citet{Shibata2016nure.book.....S}:
\begin{align}
& \partial_t \rho_* + \partial_j(\rho_* v^j) = 0, \\
& \partial_t S_i + \partial_j (S_i v^j + \alpha\sqrt{\gamma}P\delta_i{}^j) \nonumber \\
= & - S_0 \partial_i \alpha + S_j \partial_i \beta^j - \alpha \sqrt{\gamma} S_{jk}\partial_i \gamma^{jk}/2 - \alpha \sqrt{\gamma}\gamma_i{}^\mu G_\mu, \\
& \partial_t (S_0 - \rho_*) + \partial_k ((S_0 - \rho_*)v^k + \sqrt{\gamma}P (v^k + \beta^k)) \nonumber \\
= & \alpha\sqrt{\gamma}S^{ij}K_{ij} - S_i D^i \alpha + \alpha \sqrt{\gamma}n^\mu G_\mu, \\
& \partial_t(\rho_\ast Y_e) + \partial_j (\rho_\ast Y_e v^j) = - \alpha \sqrt{\gamma} \Gamma,
\end{align}
where
\begin{align}
& v^j \equiv {u^j}/{u^t}, \quad
\rho_* \equiv \alpha \sqrt{\gamma}\rho u^t, \quad
w \equiv \alpha u^t \quad \\
& S_j \equiv \rho_* h u_j, \quad
S_0 \equiv \sqrt{\gamma}(\rho h w^2 - P), \quad
S_{ij} \equiv \rho h u_i u_j + P\gamma_{ij},
\end{align}
where $\rho$, $Y_e$, $P$, $v^i$ and $u^i$ denote the density, electron fraction, pressure, 3-velocity and 4-velocity, respectively.
$G_\mu$ and $\Gamma$ denote the momentum and lepton-number feedback from neutrinos, and are calculated from the collision terms ($S_\mathrm{rad}$) as
\begin{align}
& G^\mu\equiv \sum_i \int p^\mu_i \,S_{\mathrm{rad}(i)}\, \epsilon^2 d\epsilon d(\cos\theta_\nu)d\phi_\nu, \\
& \Gamma\equiv \int (S_{\mathrm{rad}(\nu_e)}-S_{\mathrm{rad}(\bar\nu_e)})\, \epsilon^2 d\epsilon d(\cos\theta_\nu)d\phi_\nu.
\end{align}
where the index $i$ denotes the neutrino species and $p_i^\mu$ the neutrino momentum, and $\nu_e$ and $\bar\nu_e$ denote the electron-type neutrino and antineutrino, respectively.

As for the spacetime metric, the radial-gauge polar-slicing condition \citep{Gourgoulhon1991A&A...252..651G,OConnor2010CQGra..27k4103O} is imposed, and the metric is assumed to have the following form:
\begin{equation}
g_{\mu\nu} = \mathrm{diag}\left[-e^{2\Phi(t,r)},\left(1-2m(t,r)/r\right)^{-1},r^2,r^2\mathrm{sin}^2\theta\right],
\end{equation}
where the two functions $\Phi$ and $m$ are determined from the angle-averaged matter profile by solving the following equations at each time step:
\begin{equation}
\label{eq:dmdr}
\frac{\partial m}{\partial r} = 4 \pi r^2 (\rho h w^2 - P).
\end{equation}
\begin{equation}
\label{eq:dphidr}
\frac{\partial \Phi}{\partial r} = \left(1-\frac{2m}{r}\right)^{-1}
\left(\frac{m}{r^2}+4\pi r(\rho h v^2 + P)\right).
\end{equation}

\subsection{Momentum Space Grid Discretizations}
\label{sec:discretize}

The momentum space discretization method follows the one described in the Appendix of \citet{Sumiyoshi2012ApJS..199...17S}.
In the following, indices with the lowercase letter $i$ denote cell-centre values, and indices with the capital letter $I$ denote cell-interface values.

The zenith angle in momentum space, $\theta_\nu$, is gridded based on the Gauss--Legendre quadrature points for $\mu_\nu=\cos\theta_\nu$:
\begin{equation}
    \left(\mu_\nu\right)_I = \left(\mu_\nu\right)_{I-1} + \left(d\mu_\nu\right)_i.
\end{equation}
The factor $\sin^2\theta_\nu$ at the interface, which appears in the $\theta_\nu$ advection term, is evaluated as
\begin{equation}
%    \left(\sin^2\theta_\nu\right)_I =
    \left(1-\mu_\nu^2\right)_{I} = \left(1-\mu_\nu^2\right)_{I-1} - 2 \left(\mu_\nu\right)_i \left(d\mu_\nu\right)_i.
\end{equation}
This evaluation is crucial to ensure that the extra factors cancel between the different advection terms; the naive evaluation $\left(\sin^2\theta_\nu\right)_I = 1-\left(\mu_\nu\right)^2_I$ would instead leave a residual that acts as an artificial source term.

The azimuth angle is also gridded based on the Gauss--Legendre quadrature points:
\begin{equation}
    \left(\phi_\nu\right)_I = \left(\phi_\nu\right)_{I-1} + \left(d\phi_\nu\right)_i.
\end{equation}
Similarly to $\theta_\nu$, $\sin\phi_\nu$ at the interface is evaluated as
\begin{equation}
    \left(\sin\phi_\nu\right)_I = \left(\sin\phi_\nu\right)_{I-1} + \left(\cos\phi_\nu\right)_i \left(d\phi_\nu\right)_i.
\end{equation}

The energy mesh is gridded logarithmically, and the cell-centre values are determined from the cell-interface values as
\begin{equation}
    \epsilon_i = \sqrt{\epsilon_I \epsilon_{I-1}}.
\end{equation}
The integration measure for the energy integration is 
\begin{equation}
    \left(d\left(\frac{\epsilon^3}{3}\right)\right)_i = \frac{\epsilon_I^3-\epsilon_{I-1}^3}{3}.
\end{equation}
The energy mesh covers the range $\epsilon\in[0:300]\,\mathrm{MeV}$.

In this paper, the fiducial resolution is set to $N[\theta_\nu]=10$, $N[\phi_\nu]=6$ and $N[\epsilon]=20$, which is the typical resolution used in previous 1D and 2D simulations \citep{Nagakura2018ApJ...854..136N,Nagakura2019ApJ...878..160N,Nagakura2019ApJ...880L..28N,Harada2019ApJ...872..181H,Harada2020ApJ...902..150H,Akaho2023ApJ...944...60A,Akaho2024ApJ...960..116A,Akaho2025PhRvD.112d3015A,Akaho2026PhRvL.136s1002A,Akaho_2026,Barrio2025PhRvD.112h3039B}.
In order to study the resolution dependence, test cases are run in which each mesh number is individually doubled, together with lower-resolution runs for $\theta_\nu$ and $\epsilon$.

\subsection{Simulation Models}
In this paper, CCSN simulations are performed for the progenitor models with zero-age main-sequence (ZAMS) masses of $9$, $12$, $16$ and $40\,M_\odot$ taken from \citet{Sukhbold2016ApJ...821...38S}.
As for the nuclear-matter equation of state (EOS), three EOSs are employed: the Furusawa--Togashi EOS based on the variational method (VM EOS) \citep{Furusawa2017NuPhA.957..188F}, the Dirac--Br\"uckner--Hartree--Fock (DBHF EOS) \citep{Furusawa2020PTEP.2020a3D05F}, and the chiral effective field theory ($\chi$EFT EOS) \citep{Furusawa2023PrPNP.12904018F}.
As for the neutrino-matter interactions, in addition to the standard set \citep{Bruenn1985ApJS...58..771B}, nucleon bremsstrahlung, inelastic electron scattering, and electron capture on light and heavy nuclei (based on the composition calculated assuming nuclear statistical equilibrium) are included.
The spatial resolution employed in this paper is the same as in the previous studies: $N[r]=384$ covering the range $r\in[0:2500,\mathrm{km}]$ ($9\,M_\odot$) and $r\in[0:5000,\mathrm{km}]$ ($12$, $16$ and $40\,M_\odot$), with a minimum grid width of $\Delta r\sim100\,\mathrm{m}$ around the proto-neutron star (PNS) surface, and $N[\theta]=128$ for the range $\theta\in[0:\pi]$.

\section{Results}
\label{sec:results}
The angular and energy resolution dependences of the explosion dynamics are discussed in Sections \ref{sec:ang} and \ref{sec:ene}, respectively.
Section \ref{sec:BH} is devoted to the 1D BH formation simulations, where the angular and energy resolution dependences are also discussed.

\subsection{Angular Resolution Dependence of the Explosion Dynamics}
\label{sec:ang}

\begin{figure*}
    \centering
    \includegraphics[width=\linewidth]{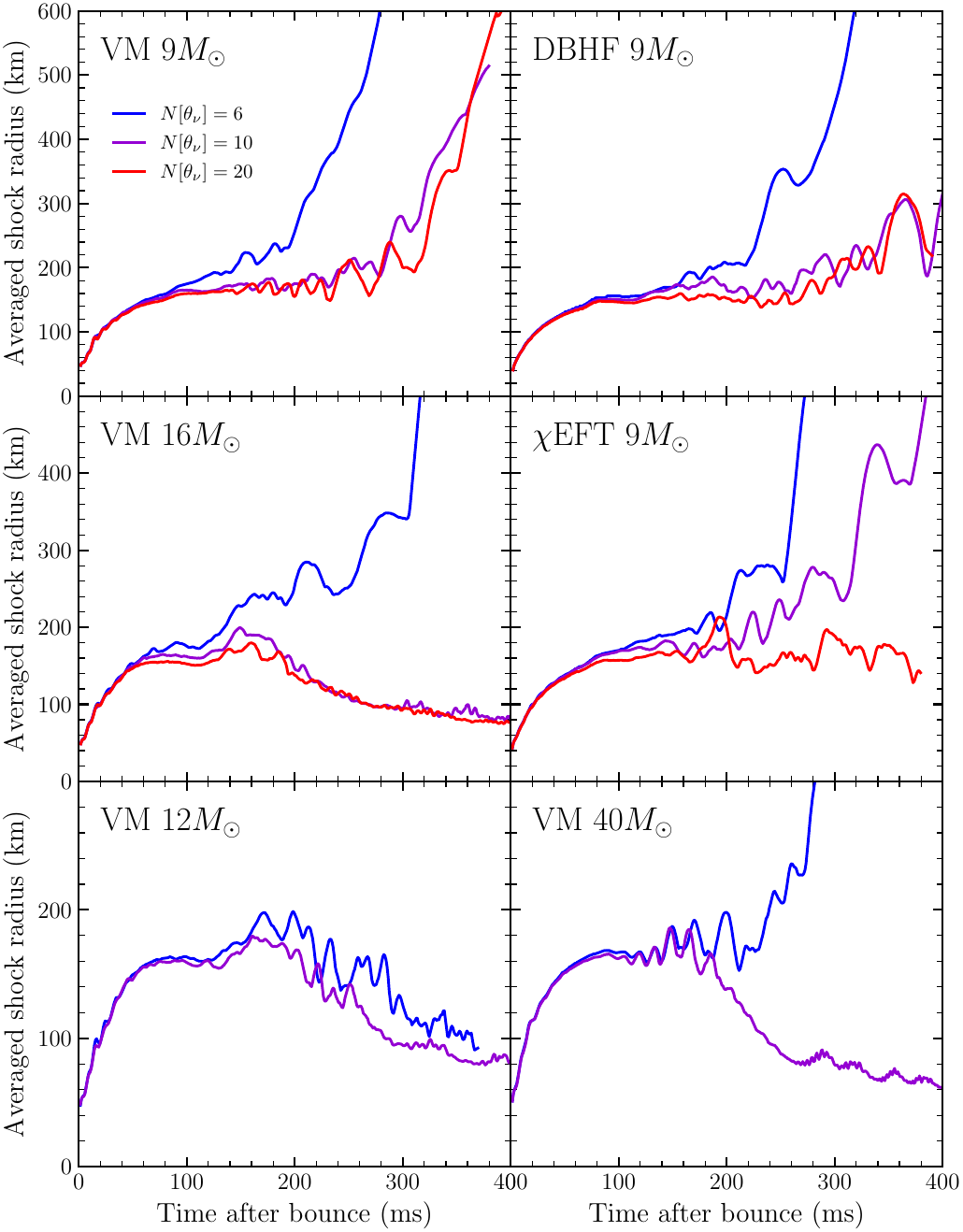}
    \caption{Time evolution of the angle-averaged shock radii. The blue, purple and red lines represent $N[\theta_\nu]=6$, $10$ (fiducial) and $20$, respectively.}
    \label{fig:shock_angdepe}
\end{figure*}
Fig. \ref{fig:shock_angdepe} shows the $\theta_\nu$ resolution dependence of the time evolution of the averaged shock radii.
For all models, the $N[\theta_\nu]=6$ case shows a more energetic and earlier shock expansion compared to the higher-resolution models.
A noteworthy feature is that the $16\,M_\odot$ and $40\,M_\odot$ progenitors, which do not explode within the simulated time at the fiducial resolution ($N[\theta_\nu]=10$), successfully explode at the low resolution $N[\theta_\nu]=6$.
At the higher resolution $N[\theta_\nu]=20$, the shock radii tend to be smaller than at the fiducial resolution, but this difference is smaller than that between the low and fiducial resolutions.
This monotonic behaviour suggests that a lower $\theta_\nu$ resolution artificially facilitates the explosion.

\begin{figure}
    \centering
    \includegraphics[width=\linewidth]{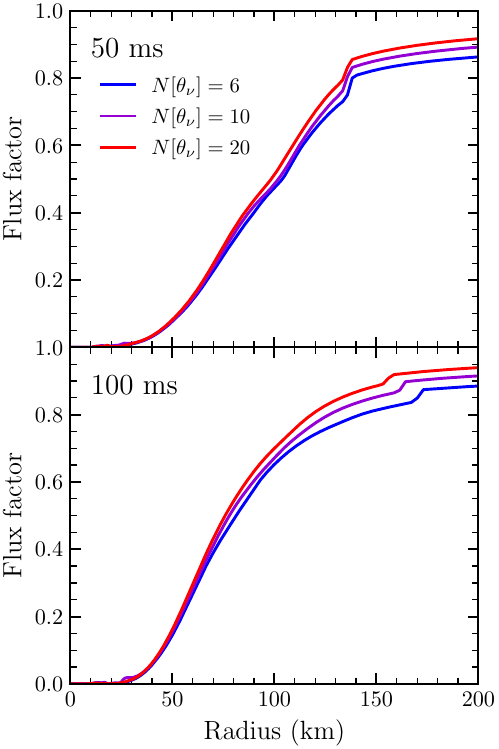}
    \caption{Radial profiles of the flux factor of $11.1\,\mathrm{MeV}$ neutrinos on the equator ($\theta=\pi/2$), for the VM EOS $9M_\odot$ model, at $50\,\mathrm{ms}$ (top) and $100\,\mathrm{ms}$ (bottom) after bounce. Different colours show different resolutions.}
    \label{fig:fluxfac_rad}
\end{figure}

This monotonic behaviour can be understood as follows.
The flux factor is defined as \citep{Shibata2011PThPh.125.1255S,Harada2019ApJ...872..181H}
\begin{equation}
\label{eq:fluxfac}
\mathscr{F} \equiv \sqrt{\frac{h_{\mu\nu}H^\mu H^\nu}{J^2}},
\end{equation}
where the projection metric onto the fluid rest frame is defined as
\begin{equation}
h_{\mu\nu} = g_{\mu\nu} + u_\mu u_\nu,
\end{equation}
and the energy density and flux in the fluid rest frame are given as
\begin{equation}
J=u_\alpha u_\beta M^{\alpha\beta},\quad H^\mu = h^\mu_\alpha u_\beta M^{\alpha\beta},
\end{equation}
where the second moment of the distribution function is
\begin{equation}
M^{\mu\nu}\equiv  \frac{1}{\epsilon} \int f {p^\prime}^\mu {p^\prime}^\nu d\Omega_p^\prime,
\end{equation}
where $d\Omega_p^\prime$ is the solid angle measure in momentum space.

Fig. \ref{fig:fluxfac_rad} shows the radial profiles of the flux factor of $11.1\,\mathrm{MeV}$ neutrinos on the equator for the VM EOS $9\,M_\odot$ model, at $50$ and $100\,\mathrm{ms}$ after bounce.
At $t=50\,\mathrm{ms}$, the shock radii are almost the same among the different models, whereas at $100\,\mathrm{ms}$ they start to deviate.
It is clear that a lower resolution leads to lower values of the flux factor.
\rev{Note that it is the flux (the numerator in Eq.~\ref{eq:fluxfac}) that varies monotonically with resolution; the angle-integrated energy density (the denominator) shows no strong resolution dependence.}
This is expected, because the decoupled neutrinos become increasingly forward-peaked as they propagate outward, and such a strongly forward-peaked distribution cannot be captured at low $\theta_\nu$ resolution.
\rev{Underestimation of the flux factor in a low angular resolution has already been pointed out in previous 1D studies \citep{Yamada1999A&A...344..533Y,Liebendorfer2004ApJS..150..263L}.}
The flux factor is an indicator of the ``average'' propagation speed of neutrinos. Therefore, neutrinos tend to stay in the gain region longer at low $\theta_\nu$ resolution, which enhances their contribution to the matter heating.

\begin{figure}
    \centering
    \includegraphics[width=\linewidth]{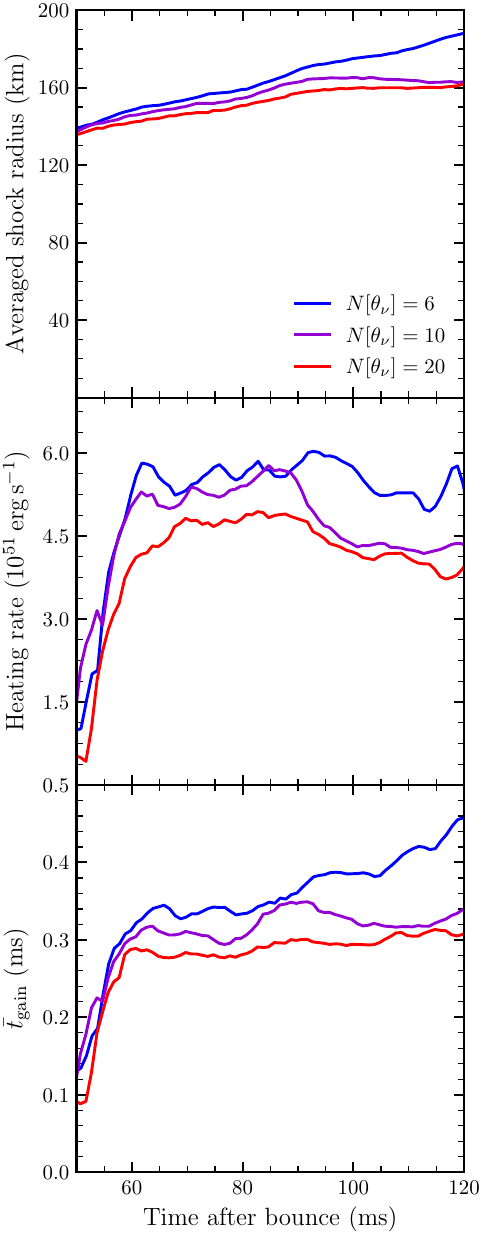}
    \caption{Time evolution of the angle-averaged shock radius (top), the total neutrino heating rate (middle) and the average neutrino stay time in the gain region $\bar t_\mathrm{gain}$ (bottom) for the VM EOS $9\,M_\odot$ model. Different colours show different $\theta_\nu$ resolutions.}
    \label{fig:staytime}
\end{figure}
The duration for which neutrinos stay in the gain region is quantitatively estimated as follows.
As a simple heuristic discussion, consider an imaginary test particle that represents the bulk of the neutrinos and propagates with speed $\mathscr{F}$.
Since its equation of motion is $\dot r(t) = \mathscr{F}(r(t))$, the ``average'' stay time of neutrinos $\bar t_\mathrm{gain}$ can be defined as the value satisfying the following equation:
\begin{equation}
    r_\mathrm{sh} - r_\mathrm{gain} = \int_{t=0}^{\bar t_\mathrm{gain}}\mathscr{F}(r(t))\,dt,
\end{equation}
where $r_\mathrm{sh}$ and $r_\mathrm{gain}$ represent the shock and gain radii, respectively. The time $t=0$ corresponds to the moment when the test particle starts propagating from the gain radius, i.e. $r(t=0)=r_\mathrm{gain}$.

Fig. \ref{fig:staytime} shows the time evolution of the shock radius, the neutrino heating rate, and $\bar t_\mathrm{gain}$.
Here, the focus is on the time range where the shock radii start to diverge.
The smaller flux factor at lower resolution leads to a longer $\bar t_\mathrm{gain}$, as naturally expected.
Since a longer stay of neutrinos in the gain region enhances the neutrino heating, the total heating rate is correlated with $\bar t_\mathrm{gain}$.
It is therefore concluded that an insufficient zenith angle resolution in momentum space artificially makes neutrinos linger in the gain region and facilitates the explosion.
\rev{The monotonic dependence of the heating rate on the angular resolution, and its association with the flux factor, was already pointed out by \citet{Messer1998}.}
Because it can even alter the explodability of a model, a coarse zenith angle resolution is a serious concern for 2D explosion simulations.
On the other hand, the fiducial resolution adopted in previous studies gives results much closer to those at the higher resolution than the lower resolution ones. Since the results are not resolution-converged, the correct answer cannot be determined yet, but the fiducial resolution results obtained so far are likely to be reasonably reliable.

\begin{figure*}
    \centering
    \includegraphics[width=0.9\linewidth]{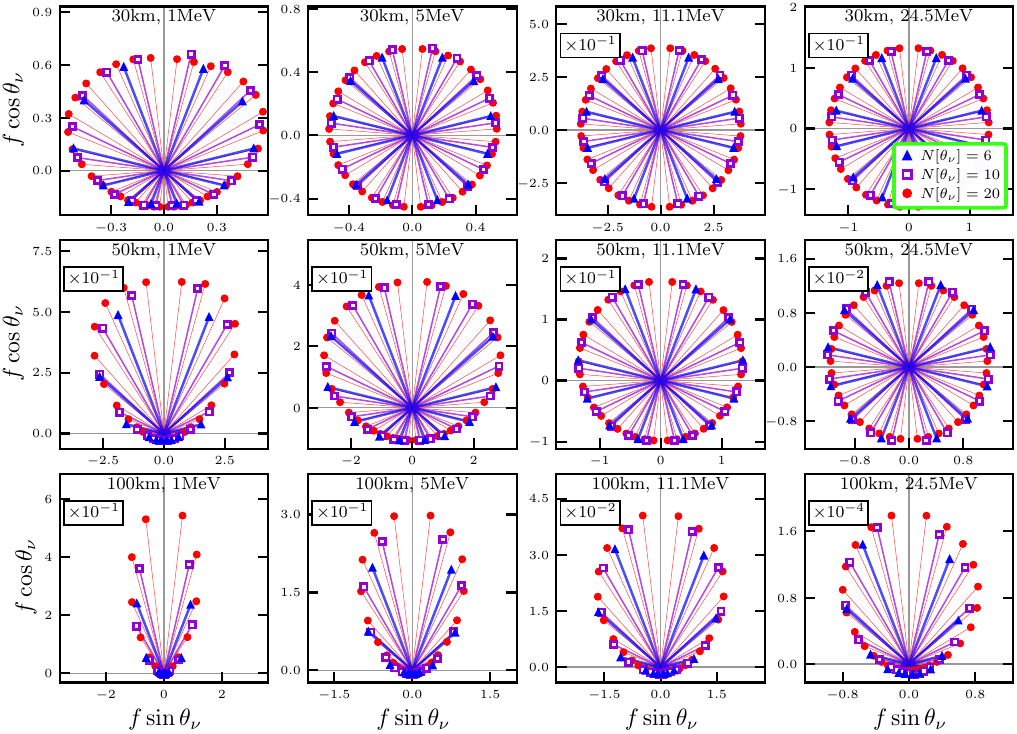}
    \caption{Zenith angle distributions on the equator for the VM EOS $9M_\odot$ model at $50\,\mathrm{ms}$ after bounce. Distributions for different radii and energies of neutrinos are shown. The right half of each panel corresponds to $\phi_\nu=0.1\pi$ and the left half corresponds to $\phi_\nu=1.1\pi$.}
    \label{fig:momspace}
\end{figure*}
For reference, the zenith angle distribution is comprehensively displayed in Fig. \ref{fig:momspace}.
As expected, the distributions are more forward-peaked at lower energies (leftward) and at larger radii (downward).
The limitation of the low-resolution model is clear: it fails to represent the most forward direction.
For example, consider the most forward-peaked case (bottom-left panel, for $100\,\mathrm{km}$ and $1\,\mathrm{MeV}$).
The distribution function in the most forward direction is $f\sim 0.2$ (at $\theta_\nu=21^\circ$) for the $N[\theta_\nu]=6$ model, whereas the most forward-peaked direction in the $N[\theta_\nu]=20$ model has $f\sim0.5$ (at $\theta_\nu=6.7^\circ$).
This clearly demonstrates that the forward-peakedness is underestimated at low $N[\theta_\nu]$.

\begin{figure}
    \centering
    \includegraphics[width=\linewidth]{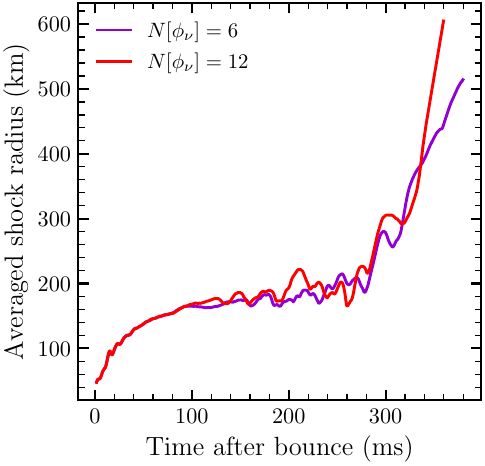}
    \caption{Time evolution of the averaged shock radii of VM EOS $9M_\odot$ model, for $N[\phi_\nu]=6$ (fiducial) and $N[\phi_\nu]=12$.}
    \label{fig:shock_phinu}
\end{figure}
Fig. \ref{fig:shock_phinu} shows the $\phi_\nu$ resolution dependence of the shock radii.
Unlike the $\theta_\nu$ dependence (Fig. \ref{fig:shock_angdepe}), the shock evolution shows no secular difference throughout the evolution.
There are fluctuations in the shock radii, but these are mainly driven by the stochasticity of the explosion.
This result is understandable because the $\phi_\nu$ resolution is unimportant while the shock remains nearly spherical, which is the case for most of the evolution.
\rev{One may then ask whether it is safe to assume complete $\phi_\nu$ symmetry (i.e. azimuthal symmetry in momentum space). A definitive answer cannot yet be given. Since the ray-by-ray approximation is known to artificially facilitate the explosion in 2D \citep{Skinner2016ApJ...831...81S,Just2018MNRAS.481.4786J} -- although this effect becomes weaker in 3D \citep{Glas2019ApJ...873...45G} -- an overly coarse $\phi_\nu$ grid may act in a similar way. Therefore, it is difficult to identify a truly safe choice for the $\phi_\nu$ resolution.}
\rev{Furthermore, the} weak dependence on the $\phi_\nu$ resolution may not apply to rapidly rotating cases, where a strong $\phi_\nu$ dependence exists \citep{Harada2019ApJ...872..181H,Iwakami2022ApJ...933...91I}.

\begin{figure*}
    \centering
    \includegraphics[width=\linewidth]{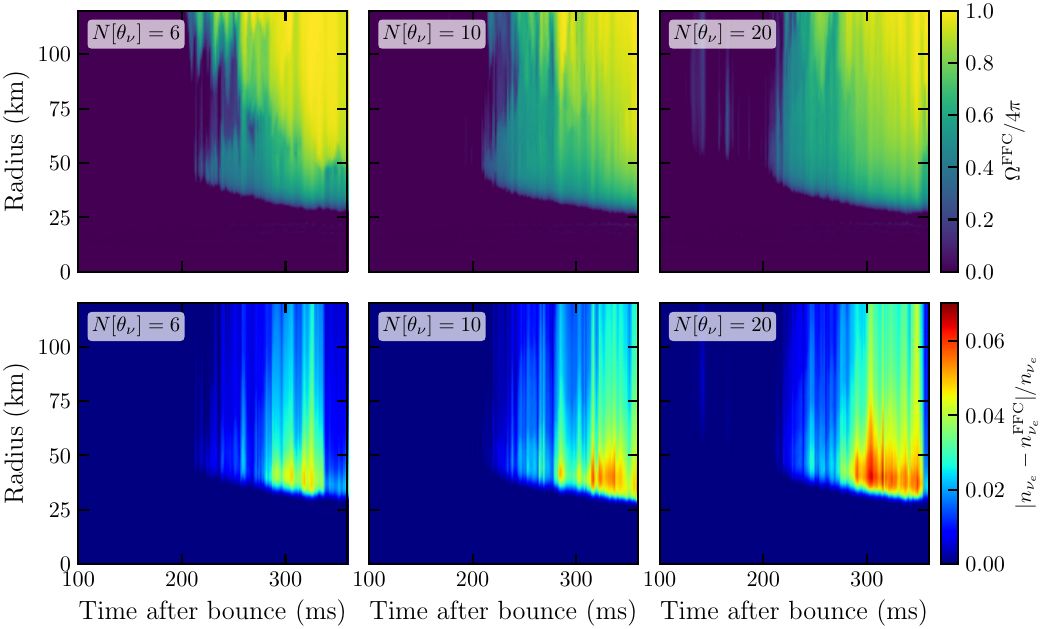}
    \caption{Time-radius map of the fraction of the solid angle with FFC $\Omega^\mathrm{FFC}/4\pi$ (top), and the relative difference of $\nu_e$ number density before and after FFC (bottom).}
    \label{fig:FFCfrac}
\end{figure*}
As already mentioned, an advantage of performing Boltzmann neutrino transport simulations is that the occurrence of FFC and the subsequent angular distributions can be treated self-consistently.
Since FFC tends to appear in the neutrino decoupling region, where neutrinos become forward-peaked \citep{Akaho2024PhRvD.109b3012A}, it is natural to expect that its appearance also depends on the angular resolution.
The fraction of the solid angle where FFI is present, $\Omega^\mathrm{FFC}/4\pi$, is shown in the top panel of Fig. \ref{fig:FFCfrac}. 
\rev{The presence of FFI is judged from the ELN crossing, i.e. from whether the quantity
\begin{equation}
\Delta G\equiv \int(f_e-\bar f_e)\,\epsilon^2\,d\epsilon,
\end{equation}
takes both positive and negative values for different momentum angles. Here $\bar\nu_e$, which decouples from the matter earlier, becomes more forward-peaked than $\nu_e$. This makes $\Delta G$ negative in the radially outward direction while it remains positive elsewhere, thereby producing the crossing. The angular distribution of $\bar\nu_e$ therefore plays an essential role in generating the FFI discussed here. Note that the contribution from the heavy-lepton type number (XLN) is ignored, since this is a post-process analysis assuming three species with $\nu_x=\bar\nu_x$.}
The FFI region starts to appear at $\sim 50\,\mathrm{ms}$ and spreads over time.
The overall behaviour is similar across resolutions, but in the lower-resolution cases there is a gap region without FFI at $\sim 75\,\mathrm{km}$ that is absent in the $N[\theta_\nu]=20$ case. This gap is therefore likely an artefact of the insufficient resolution.

The bottom panels of Fig. \ref{fig:FFCfrac} show the ``conversion rate'' of FFC, defined as the relative difference of the number density before and after FFC, $|n_{\nu_e}-n_{\nu_e}^\mathrm{FFC}|/n_{\nu_e}$.
Here $n_{\nu_e}$ denotes the current $\nu_e$ number density and $n^\mathrm{FFC}_{\nu_e}$ denotes the $\nu_e$ number density in the asymptotic state of FFC, which is determined using the method proposed by \citet{Zaizen2023PhRvD.107j3022Z}.
As shown in the figure, the relative difference increases with increasing angular resolution.
This is natural because the change of the distribution due to FFC is determined by the depth of the ELN-XLN crossing, which tends to be deeper for a more forward-peaked angular distribution.
As already seen in Fig. \ref{fig:momspace}, the forward-peakedness is underestimated at low resolution, so the crossing depth is also expected to be underestimated there.
The previous study that probed the role of FFC in CCSN simulations \citep{Akaho2026PhRvL.136s1002A} was performed with $N[\theta_\nu]=10$.
Its qualitative picture is unlikely to change with resolution, but as Fig. \ref{fig:FFCfrac} suggests, the effects of FFC would have been underestimated there.
The conclusions of \citet{Akaho2026PhRvL.136s1002A} may therefore be rather conservative.

\subsection{Energy Resolution Dependence of the Explosion Dynamics}
\label{sec:ene}

\begin{figure}
    \centering
    \includegraphics[width=\linewidth]{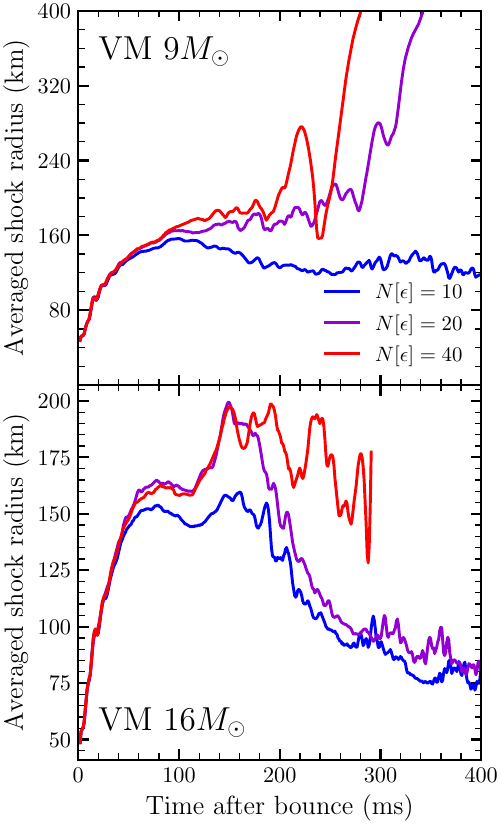}
    \caption{Time evolution of the angle-averaged shock radii for the VM EOS $9M_\odot$ (top) and VM EOS $16M_\odot$ (bottom).}
    \label{fig:shock_enedepe}
\end{figure}
Fig. \ref{fig:shock_enedepe} shows the energy resolution dependence of the shock radii.
The shock radii are greatly suppressed at the lowest resolution, $N[\epsilon]=10$.
In particular, the $9\,M_\odot$ progenitor, which explodes at the fiducial resolution, shows no sign of explosion even at $400\,\mathrm{ms}$ at the lowest resolution.
It is also observed that the higher resolution $N[\epsilon]=40$ facilitates the explosion.
For the $9\,M_\odot$ progenitor, the higher resolution leads to an earlier explosion than in the fiducial model.
The difference is less prominent for the $16\,M_\odot$ progenitor, but a higher energy resolution again leads to a more expanded shock. The $N[\epsilon]=20$ and $40$ models evolve similarly until the arrival of the Si/O interface at $\sim120\,\mathrm{ms}$, which triggers a divergence between them: the higher-resolution $N[\epsilon]=40$ model retains a larger shock radius thereafter.
These results suggest that a lower energy resolution artificially suppresses the explosion.
The behaviour is monotonic with resolution, in a similar manner to the $\theta_\nu$ dependence but in the opposite direction.

The reason for the energy resolution dependence is discussed below.
To simplify the situation, steady-state neutrino simulations are performed on the same fixed hydrodynamics background.
This helps to disentangle the effect of interest from the stochasticity of the hydrodynamics.
The VM $9\,M_\odot$ fiducial-resolution model at $150\,\mathrm{ms}$ after bounce is employed as the background, and steady-state pure neutrino transport simulations are performed on it with $N[\epsilon]=10$, $16$, $20$, $30$ and $40$.
One may also wonder how the gravitational redshift is affected by the resolution. For that purpose, steady-state simulations with a flat metric imposed are also run with $N[\epsilon]=10$, $20$ and $40$.

\begin{figure}
    \centering
    \includegraphics[width=\linewidth]{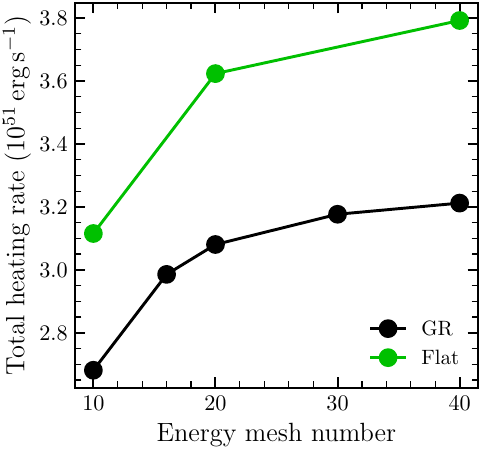}
    \caption{The total neutrino heating rate for different energy resolutions. General relativistic (GR) and flat metric cases are shown.}
    \label{fig:enedepe_mesh}
\end{figure}
Fig. \ref{fig:enedepe_mesh} shows the energy resolution dependence of the total neutrino heating rate.
The total heating rate increases monotonically with the energy mesh number, and the decreasing slope at higher resolution indicates a trend towards convergence.
The flat-metric case also shows a monotonic behaviour, with a similar degree of variation.
Therefore, the existence of the gravitational redshift itself does not strongly affect the energy resolution dependence.

\begin{figure}
    \centering
    \includegraphics[width=\linewidth]{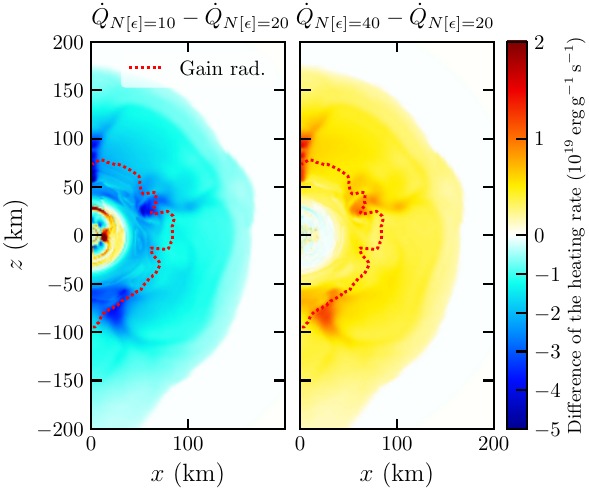}
    \caption{Meridian map of the difference of the neutrino heating rate for $N[\epsilon]=10$ and $40$ with respect to $N[\epsilon]=20$. The gain radius for $N[\epsilon]=20$ is also shown with the red dashed lines.}
    \label{fig:meridianmap_heat}
\end{figure}
Fig. \ref{fig:meridianmap_heat} shows the neutrino heating rates from the steady-state simulations for $N[\epsilon]=10$ and $40$, with the value for $N[\epsilon]=20$ subtracted. The reddish and bluish regions correspond to positive and negative values, respectively.
It is clear that the heating rate is smaller for $N[\epsilon]=10$ over a wide range of the post-shock region, both inside and outside the gain radius, while the $N[\epsilon]=40$ case clearly shows the opposite trend.
This increased (decreased) heating rate is likely responsible for the facilitated (suppressed) explosion at the higher (lower) energy resolution.

\begin{figure}
    \centering
    \includegraphics[width=\linewidth]{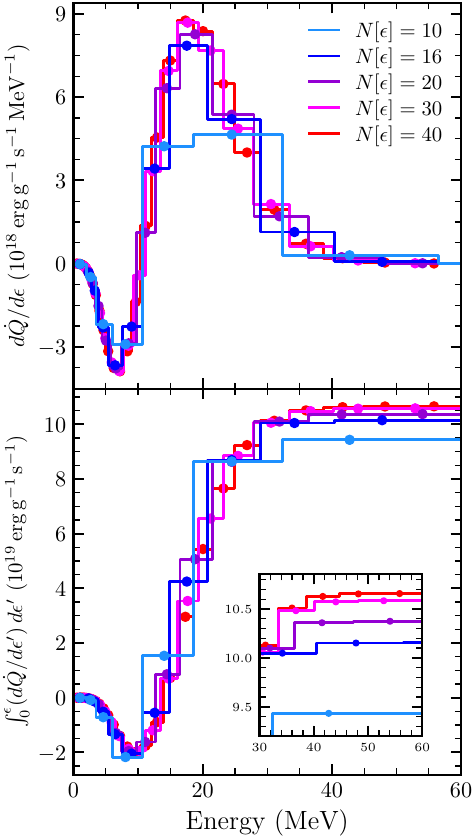}
    \caption{Differential heating rate (top) and the cumulative heating rate $\dot Q_\mathrm{cum}(\epsilon)$ (bottom), at $r=110\,\mathrm{km}$ on the equator (inside the gain region). Dots denote the cell centre values.}
    \label{fig:heatrate_ene}
\end{figure}
The next question is why the heating rate exhibits a monotonic energy resolution dependence.
Fig. \ref{fig:heatrate_ene} shows the energy-dependent (differential) neutrino heating rate at $r=110\,\mathrm{km}$ (inside the gain region) on the equator.
The low-energy neutrinos give a cooling contribution and the high-energy ones give a heating contribution.
In addition, to make the contribution of each energy bin clear, the cumulative heating rate is plotted in the bottom panel of Fig. \ref{fig:heatrate_ene}. It is the heating rate integrated up to a given energy.
The differential heating rate peaks at $\sim20\,\mathrm{MeV}$, and the $N[\epsilon]=10$ model clearly has insufficient grid points to represent such a structure.
The higher-resolution models have rather consistent values, and their cumulative heating rates almost agree up to $\sim30\,\mathrm{MeV}$. As shown in the inset, the monotonic systematic difference instead arises at higher energies.
The surprising feature is the monotonic ordering at $\gtrsim30\,\mathrm{MeV}$, where the heating rate is higher for higher energy resolution, without any exception.
Because the energy mesh is gridded logarithmically, the grid points become sparse at high energies, so the heating contribution there is underestimated when the resolution is coarse.
This clearly suggests that the lack of resolution in the intermediate energy region, $\sim30$--$50\,\mathrm{MeV}$, affects the explosion dynamics.
Importantly, whereas the angular resolution is specific to multi-angle schemes such as the Boltzmann method, the energy discretization is common to the moment method as well. Therefore, this energy resolution dependence also has implications for moment-based simulations.
Finally, note that the resolution dependence seen here depends on the energy gridding method.
In this study, a logarithmically gridded energy mesh is employed. With a different gridding method, the situation may differ.
\rev{A different aspect of the energy resolution was discussed by \citet{Mezzacappa1993ApJ...405..637M}, who found that a coarse energy mesh placed near the Fermi surface can cause unstable oscillations of the solution.}

\subsection{Resolution Dependence of the Black Hole Formation Dynamics}
\label{sec:BH}

\begin{figure}
    \centering
    \includegraphics[width=\linewidth]{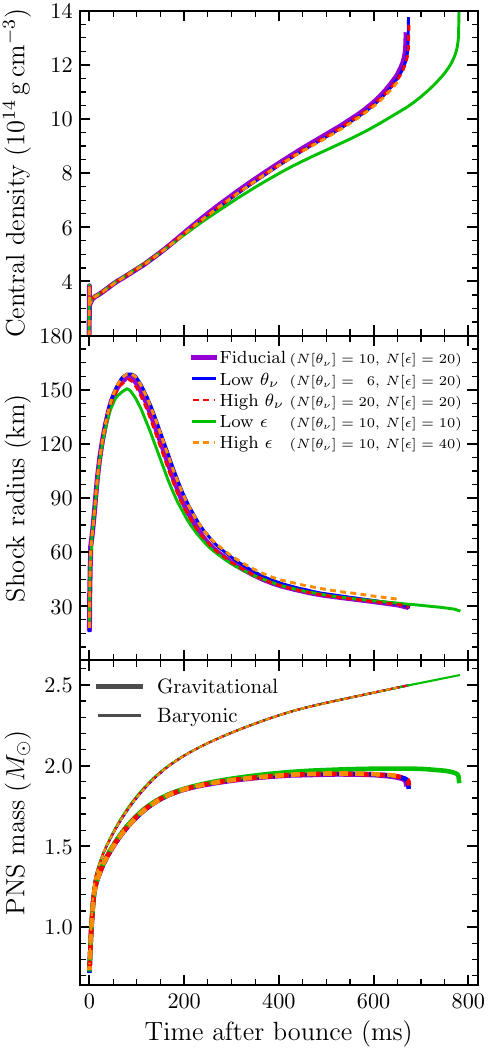}
    \caption{Time evolution of the central density (top), shock radius (middle) and neutron star mass (bottom) for the 1D $40\,M_\odot$ collapse simulations. Different colours show different resolutions.}
    \label{fig:centdensity_s40}
\end{figure}
Black hole formation from failed CCSNe, and its dependence on the progenitor structure and the nuclear EOS, has been investigated by a number of authors \citep{OConnor2011ApJ...730...70O,Pan2018ApJ...857...13P,Chan2018ApJ...852L..19C,Schneider2020ApJ...894....4D,Walk2020PhRvD.101l3013W,Burrows2023ApJ...957...68B,Janka2024Ap&SS.369...80J}.
In this section, the angular and energy resolution dependence of the 1D BH formation simulations, computed with the VM (Furusawa--Togashi) EOS, is discussed.
Fig. \ref{fig:centdensity_s40} shows the time evolution of the central density, shock radius and neutron star mass for the 1D $40\,M_\odot$ collapse simulations.
All models show a monotonically increasing central density and a jump at the final point.
This final jump corresponds to the gravitational collapse when the neutron star has reached its maximum mass.
The angular dependence is surprisingly small: the BH formation times agree to within $\lesssim10\,\mathrm{ms}$.
The shock radii also show little difference, with the $N[\theta_\nu]=6$ model exhibiting only a very slightly inflated shock radius.
This is in contrast with the enhanced explosion in 2D, presented in Section \ref{sec:ang}.
The reason is that the volume and radial extent of the gain region are small in 1D, so the delayed escape of neutrinos discussed in Section \ref{sec:ang} does not have a strong impact on the dynamics.

The mesh number $N[\theta_\nu]=6$ has been rather widely used in previous 1D simulations, including BH formation studies
\citep{Mezzacappa2001PhRvL..86.1935M,Liebendorfer2001PhRvD..63j3004L,Liebendorfer2005ApJ...620..840L,Sumiyoshi2005ApJ...629..922S,Sumiyoshi2006PhRvL..97i1101S,Sumiyoshi2007ApJ...667..382S,Sumiyoshi2008ApJ...688.1176S,Sumiyoshi2009ApJ...690L..43S,Sumiyoshi2019ApJ...887..110S,Fischer2020PhRvD.102l3001F,Fischer2020PhRvC.102e5807F,Fischer2021PhRvD.104j3012F,Xiong2024PhRvD.109l3008X,Xiong2025PhRvD.112f3024X}.
In contrast to the 2D explosion dynamics, where such a coarse zenith angle grid seriously affects the outcome (Sec. \ref{sec:ang}), the current results suggest that $N[\theta_\nu]=6$ does not pose a serious resolution problem for 1D simulations.

In contrast to the angular resolution dependence, the energy resolution has a noticeable impact.
The low-resolution model, $N[\epsilon]=10$, shows a delayed BH formation time.
The fiducial and higher-resolution ($N[\epsilon]=40$) models show almost the same BH formation time, which suggests that the result is nearly converged with respect to the energy resolution.
This energy dependence is not as severe as in the explosion simulations discussed in Section \ref{sec:ene}.

A remarkable feature seen in the bottom panel of Fig. \ref{fig:centdensity_s40} is that a lower resolution can support a neutron star with larger gravitational mass.
\begin{figure}
    \centering
    \includegraphics[width=\linewidth]{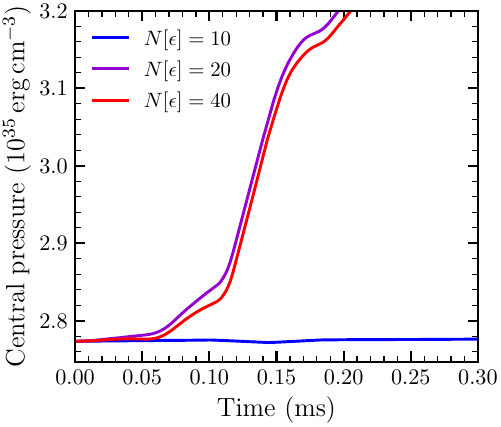}
    \caption{Time evolution of the central pressure, for the simulations started from the maximum gravitational mass NS for $N[\epsilon]=10$ model.}
    \label{fig:cent_maxNS}
\end{figure}
To investigate this difference, additional simulations are performed.
Taking the maximum-gravitational-mass configuration from the lowest-resolution model ($1.983\,M_\odot$, at $643\,\mathrm{ms}$ after bounce), higher-resolution runs are continued from that profile.
Fig. \ref{fig:cent_maxNS} shows the resulting time evolution of the central pressure.
The line for the lowest-resolution model, $N[\epsilon]=10$, is plotted for reference; it is almost constant because the profile is quasi-steady with that resolution.
The higher-resolution simulations show a rapidly increasing central pressure. This indicates that the maximum-gravitational-mass NS achieved with $N[\epsilon]=10$ is indeed unstable at higher resolution.
In other words, a lower resolution can artificially support a more massive NS.
\begin{figure}
    \centering
    \includegraphics[width=\linewidth]{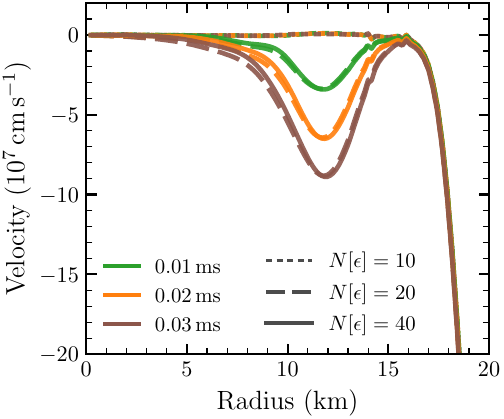}
    \caption{Radial profile of the radial velocity for the maximum mass NS test.}
    \label{fig:deform_maxNS}
\end{figure}
Fig. \ref{fig:deform_maxNS} shows the radial profile of the radial velocity after the maximum-mass NS test, in order to identify where the deformation occurs.
It is clear that the higher-resolution models show a rapid inward acceleration of the velocity at $\sim12\,\mathrm{km}$, which is the cause of the rapid shrinkage.

\begin{figure}
    \centering
    \includegraphics[width=\linewidth]{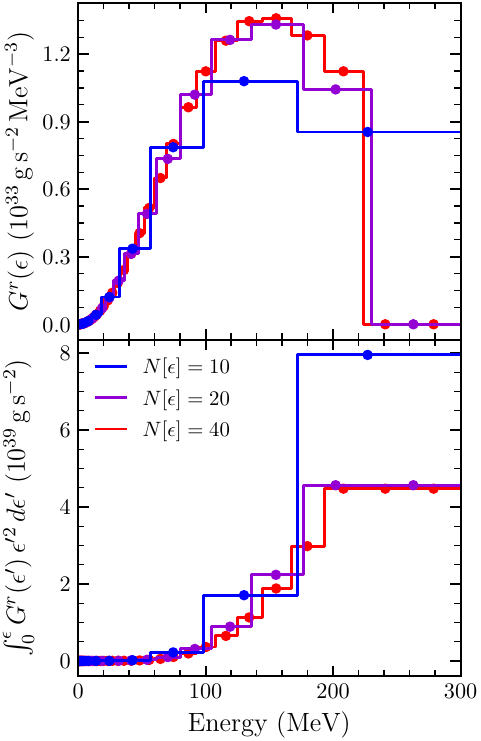}
    \caption{Energy dependence of the neutrino's momentum feedback onto matter (top), and its cumulative value (bottom).}
    \label{fig:momfeedback}
\end{figure}
The energy resolution dependence in the NS test is in fact due to the difference in the neutrino momentum feedback onto the matter.
The deformation region seen in Fig. \ref{fig:deform_maxNS} corresponds to the neutrino decoupling region, where neutrinos start to become forward-peaked and provide outward momentum to the matter.
Fig. \ref{fig:momfeedback} shows the energy-dependent neutrino momentum feedback onto matter (top) and its cumulative value (bottom).
It is clear that the lack of resolution at high energies leads to an overestimation of the momentum feedback.
This can be understood from the behaviour at the high-energy end: the fiducial and higher-resolution models show a decreasing differential momentum feedback there, indicating a trend towards convergence, whereas the low-resolution model retains an almost constant value up to the highest energy. As a result, its cumulative momentum feedback becomes too large.
Physically, at high energies the exponential fall-off of the distribution function competes with the positive power of energy in the momentum-feedback integrand. At sufficiently high energy the exponential fall-off wins, so the differential feedback turns over and the cumulative value should approach a plateau. At $N[\epsilon]=10$, however, the logarithmic grid becomes too sparse at high energies to resolve the onset of this fall-off. Hence the differential feedback stays artificially flat there and, weighted by the wide high-energy bins, the cumulative value is overestimated.
For the resolution dependence discussed in Section \ref{sec:ene}, the difference instead arose from the lack of resolution at the intermediate energies, $30$--$50\,\mathrm{MeV}$.
It is subtle that a lack of energy resolution affects the dynamics differently depending on the energy region.

\section{Conclusions}
\label{sec:conclusion}
In this paper, the momentum space resolution dependence of Boltzmann neutrino radiation hydrodynamics simulations of CCSNe is studied.
The dependence on the resolutions of the zenith and azimuth angles in momentum space, and of the energy, is discussed.

First, the resolution dependence of the explosion dynamics is discussed using a series of 2D simulations.
It is found that an insufficient zenith angle resolution artificially makes neutrinos stay in the gain region for a longer time, which artificially facilitates the neutrino heating and the explosion.
It is also found that the appearance of FFI and the conversion rate of FFC are underestimated at low zenith angle resolution.
The azimuth angle resolution is found to have little effect for the present non-rotating models.
A coarse energy resolution is found to lead to an underestimation of the neutrino heating rate and to artificially suppress the explosion.
Since the zenith angle and energy resolutions bias the explosion in opposite directions, it remains unclear whether the fiducial-resolution results are biased towards or against explosion.

Finally, the angular and energy resolution dependence is discussed for the 1D BH formation model.
In contrast to the explosion simulations, BH formation does not show a strong angular dependence.
A coarse energy resolution is found to artificially delay the BH formation time, because the artificially enhanced momentum feedback allows the lower-resolution model to support a more massive NS.

This paper has demonstrated that the results are not quantitatively converged with respect to the momentum space resolution.
Moreover, whereas the angular resolution dependence is specific to multi-angle transport, the energy resolution dependence is shared by the moment method, so the present energy-resolution results are relevant to the widely used moment-based CCSN simulations as well.
Even with the recent growth in computational power, Boltzmann neutrino transport simulations of CCSNe remain very computationally demanding, and resolution-converged simulations may still be out of reach.
Raising the resolution without raising the computational cost may require some ingenuity (e.g. treating only the advection term at high resolution; \citealt{Ito2026ApJS..284...11I}).
In that respect, this study provides useful insight for interpreting simulation results obtained with the Boltzmann transport.

\section*{Acknowledgements}
The author thanks Hiroki Nagakura for fruitful discussions. 
This work used high performance computing resources provided by Fugaku supercomputer at RIKEN, the Wisteria provided by JCAHPC through the HPCI System Research Project (Project ID: 230056, 230204, 230270, 240041, 240079, 240219, 240264, 250006, 250166, 250191, 250226, 250326).
This work is supported by JSPS KAKENHI Grant Number 24K00632, 26K17158.

%%%%%%%%%%%%%%%%%%%%%%%%%%%%%%%%%%%%%%%%%%%%%%%%%%
\section*{Data Availability}
The data used for this article will be shared on reasonable request to the corresponding author.

%%%%%%%%%%%%%%%%%%%% REFERENCES %%%%%%%%%%%%%%%%%%

% The best way to enter references is to use BibTeX:

\bibliographystyle{mnras}
\bibliography{example} % if your bibtex file is called example.bib

% Alternatively you could enter them by hand, like this:
% This method is tedious and prone to error if you have lots of references
%\begin{thebibliography}{99}
%\bibitem[\protect\citeauthoryear{Author}{2012}]{Author2012}
%Author A.~N., 2013, Journal of Improbable Astronomy, 1, 1
%\bibitem[\protect\citeauthoryear{Others}{2013}]{Others2013}
%Others S., 2012, Journal of Interesting Stuff, 17, 198
%\end{thebibliography}

%%%%%%%%%%%%%%%%%%%%%%%%%%%%%%%%%%%%%%%%%%%%%%%%%%

%%%%%%%%%%%%%%%%% APPENDICES %%%%%%%%%%%%%%%%%%%%%

%\appendix
%\section{Some extra material}

%%%%%%%%%%%%%%%%%%%%%%%%%%%%%%%%%%%%%%%%%%%%%%%%%%

% Don't change these lines
\bsp	% typesetting comment
\label{lastpage}
\end{document}